\documentclass[a4paper,11pt]{article}
\usepackage{amsthm,amssymb}
\usepackage{amsbsy,amsfonts,amsmath,amscd}
\usepackage{mathrsfs}

\usepackage{cite}
\usepackage{array}
\usepackage{epsfig,multicol}
\usepackage{verbatim}

\let\oldPhi=\Phi
\let\oldPsi=\Psi
\let\oldGamma=\Gamma
\let\oldDelta=\Delta
\let\oldSigma=\Sigma
\let\oldLambda=\Lambda
\let\oldTheta=\Theta
\let\oldPi=\Pi
\renewcommand{\Phi}{\mathnormal{\oldPhi}}
\renewcommand{\Psi}{\mathnormal{\oldPsi}}
\renewcommand{\Gamma}{\mathnormal{\oldGamma}}
\renewcommand{\Sigma}{\mathnormal{\oldSigma}}
\renewcommand{\Delta}{\mathnormal{\oldDelta}}
\renewcommand{\Theta}{\mathnormal{\oldTheta}}
\renewcommand{\Lambda}{\mathnormal{\oldLambda}}
\renewcommand{\Pi}{\mathnormal{\oldPi}}

\newcommand{\gen}[1]{\mathfrak{#1}}
\newcommand{\genY}[1]{\widehat{\mathfrak{#1}}}
\newcommand{\cwgen}[1]{\mathfrak{#1}}
\newcommand{\csgen}[1]{\mathfrak{#1}}

\newcommand{\rmat}{\mathcal{R}}

\newcommand{\crmat}{r}

\newcommand{\perm}{\mathcal{P}}

\newcommand{\psucentral}{\mathfrak{psu}(2|2)\ltimes\mathbb{R}^3}

\newcommand{\str}{\mathop{\mathrm{str}}}

\newcommand{\cartq}[1]{A_{#1}(q)}

\ifx\genfrac\sdflkaj

\else

\fi
\newcommand{\sfrac}[2]{{\textstyle\frac{#1}{#2}}}
\newcommand{\half}{\sfrac{1}{2}}

\newcommand{\sprod}[2]{\left(#1,#2\right)}

\newcommand{\comm}[2]{[#1,#2]}
\newcommand{\adj}[3]{\left(\text{ad}_{#1}\right)^{#2}(#3)}

\newcommand{\scomm}[2]{[#1,#2\}}
\newcommand{\acomm}[2]{\{#1,#2\}}
\newcommand{\sacomm}[2]{\{#1,#2]}

\newcommand{\alg}[1]{\mathfrak{#1}}

\newcommand{\gltt}{\alg{gl}(2|2)}
\newcommand{\utt}{\alg{u}(2|2)}
\newcommand{\Ya}[1]{\mathcal{Y}\left( #1\right )}
\newcommand{\DY}[1]{\mathcal{DY}\left( #1\right )}

\newcommand{\nln}{\nonumber\\}

\newcommand{\earel}[1]{\mathrel{}&\hspace{-2\arraycolsep}#1\hspace{-2\arraycolsep}&\mathrel{}}
\newcommand{\eq}{\earel{=}}
\newcommand{\beq}{\begin{equation}}
\newcommand{\eeq}{\end{equation}}

\def\[{\begin{equation}}
\def\]{\end{equation}}
\def\<{\begin{eqnarray}}
\def\>{\end{eqnarray}}

\makeatletter
\def\mr@ignsp#1 {\ifx\:#1\@empty\else #1\expandafter\mr@ignsp\fi}%
\newcommand{\multiref}[1]{\begingroup%\let\protect\string%
\xdef\mr@no@sparg{\expandafter\mr@ignsp#1 \: }%
\def\mr@comma{}%
\@for\mr@refs:=\mr@no@sparg\do{\mr@comma\def\mr@comma{,}\ref{\mr@refs}}%
\endgroup}
\makeatother

\newcommand{\hypref}[2]{\ifx\href\asklfhas #2\else\href{#1}{#2}\fi}

\renewcommand{\eqref}[1]{(\multiref{#1})}

\ifx\href\asklfhas\newcommand{\href}[2]{#2}\fi

\begin{document}

\baselineskip=16pt plus 0.2pt minus 0.1pt
\begin{titlepage}
\begin{flushright}
Imperial/TP/08/FS/02\\
HU-EP-08/xx\\
\end{flushright}
\mbox{ }  \hfill
\vspace{5ex}
\Large
\begin {center}

{\bf Weakly coupled $\mathcal N=4$ Super Yang-Mills and $\mathcal N=6$ Chern-Simons theories from $\alg{u(2|2)}$ Yangian symmetry}

\end {center}
\large
\vspace{1ex}
\begin{center}
Fabian Spill
\end{center}
\vspace{1ex}
\begin{center}
\textit{Blackett Laboratory,
Imperial College London\\
London SW7 2AZ, UK\\
and\\
Humboldt-Universit\"at zu Berlin, Institut f\"ur Physik,\\%
Newtonstra\ss{}e 15, D-12489 Berlin, Germany}\vspace{3mm}\\
\texttt{fabian.spill@imperial.ac.uk}\\

\end{center}
\vspace{4ex}
\rm

\begin{center}
{\bf Abstract}
\end{center}

In this paper we derive the universal R-matrix for the Yangian $\mathcal Y(\utt)$, which is an abstract algebraic object leading to rational solutions of the Yang-Baxter equation on representations. We find that on the fundamental representation the universal R-matrix reduces to the standard rational R-matrix $\rmat =\rmat_0\left(1 + \frac{1}{u}\perm\right)$, where the scalar prefactor is surprisingly simple compared to prefactors one finds e.g. for $\alg{sl}(n)$ R-matrices. This leads precisely to the S-matrix giving the Bethe Ansatz of one-loop $\mathcal N = 4$ Super Yang-Mills theory and two-loop $\mathcal N = 6$ Chern-Simons theory. 

\normalsize

\vfill
\end{titlepage}

\section{Introduction}

Following the work \cite{Minahan:2002ve,Beisert:2003tq, Beisert:2003yb} much progress has been done the last couple of years in understanding the duality between $\mathcal N = 4$ Super Yang-Mills theory and IIB string theory on $AdS_5\times S^5$ in the large $N$ limit by using integrability. Similar techniques have been used before in the context of QCD \cite{Lipatov:1993yb,Lipatov:1994xy}. This development culminated in the proposal of long-range asymptotic Bethe equations \cite{Beisert:2005fw}, which, when suplemented with the right dressing factor \cite{Beisert:2006ib,Beisert:2006ez}, describe the full asymptotic spectrum of the theory. This Bethe ansatz can be rigorously derived by diagonalising the asymptotic S-matrix \cite{Beisert:2005tm}, whose weak coupling limit we will investigate in this paper. It allows for highly nontrivial tests of the AdS/CFT correspondence as performed in \cite{Beisert:2006ez ,Freyhult:2007pz}. Interestingly, basically the same S-matrix was argued \cite{Ahn:2008aa} to lead to Bethe equations \cite{Gromov:2008qe} conjectured to describe the spectrum of $\mathcal N = 6$ Chern-Simons theory and string theory on $AdS_4\times \mathbb{CP}^3$, a recently proposed example of an AdS/CFT correspondence \cite{Aharony:2008ug}.

The uncovering of symmetries of a physical system is of fundamental importance as symmetries constrain the dynamics and can hence help to find solutions of dynamical equations. In an integrable system, symmetries are even more important as they allow for an exact solvability of the system. As string theory and $\mathcal N = 4$ SYM/$\mathcal N = 6$ CS theories have infinitely many degrees of freedom, the underlying symmetry algebra for this system should be infinite dimensional as well. Indeed, classical string theory on $AdS_5\times S^5$ has infinitely many symmetries \cite{Bena:2003wd}, and on the gauge theory side at tree level the corresponding symmetries have shown to be the Yangian of $\alg{psu}(2,2|4)$ in \cite{Dolan:2003uh, Dolan:2004ps}. Unfortunately, it is not easy to extend this symmetry to the full quantum system as would be required to prove that the system is quantum integrable, but there has been work showing that the Yangian algebra holds at a few loops at least in certain subsectors \cite{Agarwal:2004sz,Agarwal:2005ed, Zwiebel:2005er,Zwiebel:2006cb,Beisert:2007sk,Zwiebel:2008gr}.

In this work, we will follow another line of thought uncovering the residual symmetry algebra which one gets after breaking $\alg{psu}(2,2|4)$ in the Bethe Ansatz by choosing a vacuum state. Naively, this symmetry algebra is given by $\alg{u}(1)\ltimes\alg{psu}(2|2)\times\alg{psu}(2|2)\ltimes\alg{u}(1)$, which is the same as $\utt\times\utt$ after identifying the central elements and the outer automorphisms of both $\utt =\alg{u}(1)\ltimes\alg{psu}(2|2)\ltimes\alg{u}(1)$'s \cite{Beisert:2004ry}. As one has the Yangian as a symmetry enhancement of $\alg{psu}(2,2|4)$, one should expect that $\utt\times\utt$ is also promoted to the Yangian $\Ya{\utt\times\utt}$. However, it turns out that this is not quite the case. Indeed, the spectral parameters in the Bethe Ansatz of \cite{Beisert:2005fw} do not enter in the usual way depending only on differences of the rapidities of the corresponding magnons. In \cite{Beisert:2005tm} it was shown that the S-matrix leading to the Bethe Ansatz \cite{Beisert:2005fw} is not invariant under $\utt\times\utt$ but $\alg{psu}(2|2)\times\psucentral$, where $\psucentral$ is the universal central extension of $\alg{psu}(2|2)$. The additional central elements are related to the usual spectral parameter as well as to an additional braiding element \cite{Gomez:2006va, Plefka:2006ze} which twists the universal enveloping algebra of $\psucentral$. The S-matrix is now fixed up to the dressing factor on the fundamental representation (in which the magnons live) of $\psucentral$ by only commuting it with the generators of this twisted enveloping algebra. However, even though $\psucentral$ is still finite dimensional and the matrix structure of the S-matrix is fixed on the fundamental representation without referring to any infinite dimensional symmetry algebra, it was shown in \cite{Beisert:2007ds} that this S-matrix is additionally invariant under the Yangian $\Ya{\psucentral}$, and indeed for higher dimensional representations which are e.g. important for bound state S matrices one needs this additional Yangian symmetry to fix the S-matrix \cite{deLeeuw:2008dp} if one does not want to refer to the computationally more complicated Yang Baxter equation as done in \cite{Arutyunov:2008zt}. 

It would be desirable to have a universal form for the R-matrix, as it usually exists for Yangians. Then one could obtain all the matrices in the different representations by just plugging in the universal R-matrix into the representation maps. One problem is that $\psucentral$ is not simple and has no non-degenerate invariant bilinear form, which is needed to compute the quantum double and the universal R-matrix. The central charges can not be paired with anything in $\psucentral$. However, two of the central charges are some kind of gauge transformations and can be removed on the level of semi-classical strings, and the remaining central charge can be dually paired with one of the external automorphisms of $\alg{psu}(2|2)$ \cite{Beisert:2007ty}. However, this works only on the level of the loop algebra of $\utt$, which is deformed as a reminiscence of the additional central charges and the braiding not appearing explicitly on the classical level. The appearance of the loop algebra is expected as it is the classical analog of the Yangian double.

The classical r-matrix mentioned above arises as a limit $R = 1 + \frac{1}{g} r$, where $g$ is proportional to the square root of the 't Hooft coupling, it was first studied in \cite{Torrielli:2007mc}. One can also expand $R = 1 + g r$ and finds that the corresponding classical r-matrix is obtained from the undeformed loop algebra of $\utt$. In this paper we will obtain the mathematical quantisation of the quasi-triangular Lie bialgebra $\utt[u,u^{-1}]$ with its classical r-matrix $\crmat = \frac{\mathcal T}{u}$. By this we mean that we find the quasi-triangular Hopf algebra which has this Lie bialgebra as its classical limit. This quantisation is the Yangian double $\DY{\utt}$, which on the fundamental evaluation representation leads to the R-matrix $\rmat \propto 1 + \frac{1}{u}\perm$. It is of course not the full R-matrix of the AdS/CFT correspondence describing magnon scattering at weak and string excitation scattering at strong coupling; it does not depend on $g$. This is not surprising as the classical Lie bialgebra we quantise does not know about the central extension and braiding which are necessary to encode the coupling dependence in the representation of the Lie algebra. The rational R-matrix is nevertheless an interesting physical object: it describes the scattering of magnons at one-loop $\mathcal N = 4$ SYM/two loop $\mathcal N = 6$ CS theories. As this R-matrix is used to derive the Bethe Ansatz which yields the spectrum of the theories, it effectively means that almost the full dynamics at one-loop SYM/two-loop CS theory is determined by symmetries only.

We should emphasis also that the universal R matrices which have been obtained for Yangians of simple Lie algebras lead on evaluation representations to rational R matrices with quite complicated prefactors containing products of gamma functions \cite{Khoroshkin:1994uk}. When considering e.g. Heisenberg XXX spin chains one has to drop these prefactors. It is very interesting that in the $\utt$ case considered here no such prefactor appears and the R-matrix is a simple rational function of the difference of the spectral parameters only.

We will start in section \ref{sec:ygltt} by giving the definition of the Yangian $\mathcal Y(\gltt)$ in Drinfeld's second realization which is suitable for the derivation of the universal R-matrix in section \ref{sec:universalR}. The derivation follows the methods developed in \cite{Khoroshkin:1994uk} for simple Lie algebras, and we will highlight the modifications necessary for the case of $\gltt$ considered here. In section \ref{sec:funrep} we will study the fundamental representation of $\mathcal Y(\gltt)$ and evaluate the universal R-matrix on the tensor product of two fundamental representation, finding an R-matrix proportional to the standard rational R-matrix 

Let us note that as in this paper we mainly deal with pure algebra, we do not choose reality conditions and work with the complexified Lie algebra $\gltt$ of $\utt$.

\section{The Yangian $\Ya{\gltt}$}\label{sec:ygltt}

In this section we define the Yangian of $\gltt$ in Drinfeld's second realization. Let us first start by giving the definition of the simple Lie superalgebra $\alg{psl}(2|2)$, whose distinguished Cartan matrix is given by

\[
a_{\alg{psl}(2|2)} = \begin{pmatrix}
2&-1&0\\
-1&0&1\\
0&1&-2
\end{pmatrix}.
\]

The corresponding Chevalley-Serre generators $\gen{H}_i, \gen{E}^\pm_i$ satisfy the usual commutation relations

\<
\comm{\gen{H}_i}{\gen{H}_j} &=& 0, \nonumber\\
\comm{\gen{H}_i}{\gen{E}^\pm_j} &=& \pm {a_{ij}}\gen{E}_j^\pm,\nonumber  \\
\comm{\gen{E}^+_i}{\gen{E}^-_j} &=& \delta_{ij}{\gen{H}_i},
\>
and the Serre relations
\<
\adj{{\gen{E}}_1^\pm}{2}{{\gen{E}}_2^\pm} = \adj{{\gen{E}}_2^\pm}{2}{{\gen{E}}_1^\pm} = 0, \nonumber\\
\adj{{\gen{E}}_3^\pm}{2}{{\gen{E}}_2^\pm} = \adj{{\gen{E}}_2^\pm}{2}{{\gen{E}}_3^\pm} = 0, \nonumber\\
\comm{{\gen{E}}_1^\pm}{\gen{E}_3^\pm} = 0, \nonumber\\
\acomm{\comm{{\gen{E}}_2^\pm}{\gen{E}_1^\pm}}{\comm{{\gen{E}}_2^\pm}{\gen{E}_3^\pm}} = 0 .
\>

In this distinguished basis $\gen{E}_{1,3}^\pm$ are bosonic whereas $\gen{E}_{2}^\pm$ is fermionic, and we use the definition $\adj{\gen{X}}{1}{\gen{Y}} := \scomm{\gen{X}}{\gen{Y}}:= \gen{X}\gen{Y} - (-1)^{|\gen{X}||\gen{Y}|}\gen{Y}\gen{X}$, whereas $\comm{\gen{X}}{\gen{Y}} := \gen{X}\gen{Y} - \gen{Y}\gen{X}$, $\acomm{\gen{X}}{\gen{Y}} := \gen{X}\gen{Y} + \gen{Y}\gen{X}$ and $\sacomm{\gen{X}}{\gen{Y}} :=  \gen{X}\gen{Y} + (-1)^{|\gen{X}||\gen{Y}|}\gen{Y}\gen{X}$.Furthermore, we note that in $\alg{psl}(2|2)$ the light-like combination $\gen{H}_1 + 2 \gen{H}_2 + \gen{H}_3$ is zero.

From this definition it is straightforward to write down the relations for the generators $\csgen{E}^\pm_{i,n}$, $\csgen{H}_{i,n} $, $i=1,2,3$ and $n=0,1,\dots$ of the Yangian $\Ya{\alg{psl}(2|2)}$,

\begin{align}
\label{def:ypsl22}
&[\csgen{H}_{i,m},\csgen{H}_{j,n}]=0,\quad [\csgen{H}_{i,0},\csgen{E}^+_{j,m}]=a_{ij} \,\csgen{E}^+_{j,m},\nonumber\\
&[\csgen{H}_{i,0},\csgen{E}^-_{j,m}]=- a_{ij} \,\csgen{E}^-_{j,m},\quad \scomm{\csgen{E}^+_{i,m}}{\csgen{E}^-_{j,n}}=\delta_{i,j}\, \csgen{H}_{j,n+m},\nonumber\\
&[\csgen{H}_{i,m+1},\csgen{E}^\pm_{j,n}]-[\csgen{H}_{i,m},\csgen{E}^\pm_{j,n+1}] = \pm\frac{1}{2} a_{ij} \{\csgen{H}_{i,m},\csgen{E}^\pm_{j,n}\},\nonumber\\
&\scomm{\csgen{E}^\pm_{i,m+1}}{\csgen{E}^\pm_{j,n}}-\scomm{\csgen{E}^\pm_{i,m}}{\csgen{E}^\pm_{j,n+1}} = \pm\frac{1}{2} a_{ij} \sacomm{\csgen{E}^\pm_{i,m}}{\csgen{E}^\pm_{j,n}},
\end{align}
\begin{eqnarray}
&&i\neq j, \, \, \, \, \, n_{ij}=1+|a_{ij}|,\, \, \, \, \, Sym_{\{k\}} [\csgen{E}^\pm_{i,k_1},[\csgen{E}^\pm_{i,k_2},\dots [\csgen{E}^\pm_{i,k_{n_{ij}}}, \csgen{E}^\pm_{j,l}\}\dots\}\}=0,\nonumber\\
&&\acomm{\comm{{\csgen{E}}_{2,k_1}^\pm}{\csgen{E}_{1,k_2}^\pm}}{\comm{\csgen{E}_{2,k_3}^\pm}{\csgen{E}_{3,k_4}^\pm}} = 0, k_j = 0,1, \dots .
\end{eqnarray}

All these relations follow the general methods outlined for simple Lie algebras in \cite{Drinfeld:1987sy} and for $\alg{psl}(n|m), n\neq m$ in \cite{Stukopin:2005aa}. At this point there is no modification necessary compared to the general theory. The clear distinction of the case of $n = m$ is the fact that the Cartan matrix is degenerate. This is an obstruction to obtain the quantum double and the corresponding universal R-matrix. As explained in the introduction, this is also not what we really want. We will hence extend $\alg{psl}(2|2)$ to its central extension $\alg{sl}(2|2)$, and also adjoin the outer automorphism getting $\alg{gl}(2|2)$. Let us note that the way we do this extension works equally well for all $\alg{psl}(n|n)$, even though we did not calculate these cases explicitly.

$\alg{sl}(2|2)$ is obtained from $\alg{psl}(2|2)$ by relaxing the condition $\gen{H}_1 + 2 \gen{H}_2 + \gen{H}_3 = 0$ to

\[
\gen{C} := \frac{1}{2}(\gen{H}_1 + 2 \gen{H}_2 + \gen{H}_3) \qquad\text{central} .
\]

We furthermore introduce an additional Cartan generator $\gen{H}_4$ such that the extended Cartan matrix is given by

\[
a = \begin{pmatrix}
2&-1&0&0\\
-1&0&1&1\\
0&1&-2&0\\
0&1&0&0\\
\end{pmatrix}.
\]

This means that $\gen{H}_4$ is precisely the dual operator to the central charge $\gen{C}$, it acts as an external derivation on $\alg{sl}(2|2)$.
Concerning the modification of the definition of the Yangian the relations will look the same as in the standard case or the case of $\alg{psl}(2|2)$ in \eqref{def:ypsl22} by adjoining generators $\kappa_{4,n}$. 

In what follows we will also need a Cartan-Weyl basis for $\gltt$. We choose it the following way: Let $\alpha_1, \alpha_2, \alpha_3$ be the three simple roots corresponding to $\gen{E}_i^+, i=1,2,3$, then a set of all positive roots is given by

\[
\beta_1 = \alpha_1, \beta_2 = \alpha_1 + \alpha_2, \beta_3 = \alpha_1 + \alpha_2 + \alpha_3, \beta_4 = \alpha_2, \beta_5 =  \alpha_2 + \alpha_3, \beta_6 = \alpha_3 .
\]

The corresponding generators are given by

\begin{align}
\cwgen{E}^+_{\beta_2} &= \comm{\csgen{E}^+_1}{\csgen{E}^+_2}, \qquad &\cwgen{E}^-_{\beta_2} &= \comm{\csgen{E}^-_2}{\csgen{E}^-_1},\nonumber\\
\cwgen{E}^+_{\beta_3} &= \acomm{\comm{\csgen{E}^+_1}{\csgen{E}^+_2}}{\csgen{E}^+_3}, \qquad &\cwgen{E}^-_{\beta_3} &	= \acomm{\comm{\csgen{E}^-_1}{\csgen{E}^-_2}}{\csgen{E}^-_3},\nonumber\\
\cwgen{E}_{\beta_5}^+ &= \comm{\cwgen{E}_{2}^+}{\cwgen{E}_{3}^+}, \qquad &\cwgen{E}^-_{\beta_5} &= \comm{\cwgen{E}_{2}^-}{\cwgen{E}_{3}^-}.
\end{align}

When it comes to the Yangian one can basically think of the root $\xi^\pm_{i,n}$ as corresponding to the root vector $\pm\alpha_i + n\delta$, with $\delta$ being the affine root. Note however that the Yangian is a deformation of the universal enveloping algebra of the loop algebra, hence they are not precisely the same.

 We also give a third representation appearing in the literature \cite{Beisert:2005tm ,Beisert:2006qh} which defines  $\gltt$ close to the fundamental matrix representation as $4\times 4$ supermatrices. Let $\gen{R}_a^b$, $\gen{L}_\alpha^\beta$, $a,b,\alpha,\beta = 1\dots 2$ denote a linear bases of the bosonic $\alg{sl}(2)$ subalgebras, and the supercharges transform in the two dimensional fundamental/antifundamental of the $\alg{sl}(2)'s$ with the indices chosen appropriately, the full commutations relations read

\[
\begin{array}[b]{rclcrcl}
\comm{\gen{R}^a{}_b}{\gen{R}^c{}_d}\eq
\delta^c_b\gen{R}^a{}_d-\delta^a_d\gen{R}^c{}_b,
&&
\comm{\gen{L}^\alpha{}_\beta}{\gen{L}^\gamma{}_\delta}\eq
\delta^\gamma_\beta\gen{L}^\alpha{}_\delta-\delta^\alpha_\delta\gen{L}^\gamma{}_\beta,
\\[3pt]
\comm{\gen{R}^a{}_b}{\gen{Q}^\gamma{}_d}\eq
-\delta^a_d\gen{Q}^\gamma{}_b+\half \delta^a_b\gen{Q}^\gamma{}_d,
&&
\comm{\gen{L}^\alpha{}_\beta}{\gen{Q}^\gamma{}_d}\eq
+\delta^\gamma_\beta\gen{Q}^\alpha{}_d-\half \delta^\alpha_\beta\gen{Q}^\gamma{}_d,
\\[3pt]
\comm{\gen{R}^a{}_b}{\gen{S}^c{}_\delta}\eq
+\delta^c_b\gen{S}^a{}_\delta-\half \delta^a_b\gen{S}^c{}_\delta,
&&
\comm{\gen{L}^\alpha{}_\beta}{\gen{S}^c{}_\delta}\eq
-\delta^\alpha_\delta\gen{S}^c{}_\beta+\half \delta^\alpha_\beta\gen{S}^c{}_\delta,\nonumber
\end{array}\]
\<
\acomm{\gen{Q}^\alpha{}_b}{\gen{S}^c{}_\delta}\eq
\delta^c_b\gen{L}^\alpha{}_\delta +\delta^\alpha_\delta\gen{R}^c{}_b
+\delta^c_b\delta^\alpha_\delta\gen{C},
\nln
\acomm{\gen{Q}^{\alpha}{}_{b}}{\gen{Q}^{\gamma}{}_{d}}\eq
\varepsilon^{\alpha\gamma}\varepsilon_{bd}\gen{P},
\nln
\acomm{\gen{S}^{a}{}_{\beta}}{\gen{S}^{c}{}_{\delta}}\eq
\varepsilon^{ac}\varepsilon_{\beta\delta}\gen{K}.
\>

Here, $\gen{C} = \frac{1}{2}(\gen{R}^1_1 + \gen{R}^2_2 + \gen{L}^1_1 + \gen{L}^2_2)$ is the central element.

The relation to the Chevalley-Serre basis for $\gltt$ is given by identifying

\begin{align}\label{def:linbase}
\gen{H}_1 & = \gen{R}_1^1 - \gen{R}^2_2, \qquad & \gen{E}_1^+ &= \gen{R}^1_2, \qquad &\gen{E}_1^- &= \gen{R}^2_1, \nonumber\\
\gen{H}_2 & = \gen{R}_2^2 + \gen{L}^1_1 ,\qquad & \gen{E}_2^+ &= \gen{S}^2_1, \qquad &\gen{E}_2^- &= \gen{Q}^1_2,\nonumber\\
\gen{H}_3 & = -\gen{L}_1^1 + \gen{L}^2_2,\qquad & \gen{E}_3^+ &= \gen{L}^1_2, \qquad &\gen{E}_3^- &= -\gen{L}^2_1,\nonumber\\
\gen{H}_4 &= \frac{1}{4}(\gen{R}_1^1 + \gen{R}^2_2 - \gen{L}_1^1 - \gen{L}^2_2).
\end{align}

Now the fundamental matrix representation on the four dimensional graded vector space such that the first two base elements are even and the remaining two are odd is given as follows:

\begin{align}
\gen{R}^a_b &= E^a_b,\qquad \gen{L}^\alpha_\beta = E^{\alpha + 2}_{\beta + 2}, \nonumber\\
\gen{Q}^{\alpha}{}_{b} &= E^{\alpha+2}_b,\qquad \gen{S}_\alpha^b = E_{\alpha+2}^b .
\end{align}

Here, $E^i_j, i,j=1,\dots 4$ denotes the four by four-matrix with a one at the $i$th row and the $j$th column and zero otherwise.

\section{The universal R-matrix of $\gltt$}\label{sec:universalR}

In this section we will write down the universal R-matrix for the quantum double of $\Ya{\gltt}$, which is the double Yangian $\DY{\gltt}$. It is the quantisation of the Lie bialgebra $\gltt[u,u^{-1}]$ with classical r-matrix $\crmat = \frac{\mathcal T}{u}$, where $\mathcal T$ is the Casimir of $\gltt$. We note that for the derivation of the universal R-matrix outlined in \cite{Khoroshkin:1994uk} for simple Lie algebras or in \cite{Stukopin:2005aa} for $\alg{psl}(n|m), n\neq m$ the crucial ingredient is the existence of a non-degenerate invariant bilinear form on the Lie algebra. As this form exists for $\gltt$ the construction of the quantum double $\Ya{\gltt}$ works in the same way as for simple Lie algebras or $\alg{psl}(n|m), n\neq m$ even.

Let us briefly recall how the double construction worked on the classical case, for any Lie superalgebra with non-degenerate invariant bilinear form $\kappa$. One starts with the Lie algebra $\alg{g}[u]$ of polynomials in $u$ with values in the Lie superalgebra $\alg{g}$, whose generators we call $\gen{J}_n^a, n\in \mathbb N$, and the generators of degree $0$ are identified with the generators of $\alg{g}$. Then the double of $\alg{g}[u]$ is the loop algebra $\alg{g}[u,u^{-1}]$\footnote{We will ignore mathematical subtleties involving the fact that the dual of $\alg{g}[u]$ is in fact larger and involves infinite formal power series. In fact we will always allow for formal power series in both $u$ and $u^{-1}$. }. The pairing is given by

\[
(\gen{J}_n^a,\gen{J}_{m}^b) = \kappa^{ab}\delta_{n,-m-1} .
\]

The resulting canonical classical r-matrix for this double is given by

\[
\crmat = \sum_{n=0}^\infty \kappa_{ab}\gen{J}_n^a\otimes \gen{J}_{-n-1}^b = \frac{\mathcal T}{u_1-u_2} .
\]

upon identifying $\gen{J}_n^a = u^n\gen{J}_0^a$.

Now the double Yangian is a deformation of the universal enveloping algebra of $\alg{g}[u,u^{-1}]$, and as $\alg{g}[u,u^{-1}]$ is a quasi-triangular bialgebra, the double Yangian $\DY{\alg{g}}$ is a quasi-triangular Hopf Algebra \footnote{As the resulting classical or quantum r/R matrices are strictly speaking not elements of  $\alg{g}[u,u^{-1}]\otimes\alg{g}[u,u^{-1}]$ or $\DY{\alg{g}}\otimes\DY{\alg{g}}$ respectively, one should strictly not call the structures quasi-triangular but pseudotriangular \cite{Drinfeld:1986in}.}.

\subsection{The Universal R Matrix}

The universal R-matrix of $\DY{\gltt}$ has the form

\[
\rmat_+\rmat_H\rmat_- ,
\]

with

\<
\rmat_+ = \prod_{\alpha\in\Delta_+}^\rightarrow\exp(-a(\alpha)\cwgen{E}_\alpha^+\otimes\cwgen{E}_\alpha^-) ,\\
\rmat_- = \prod_{\alpha\in\Delta_+}^\leftarrow\exp(-a(\alpha)\cwgen{E}_\alpha^-\otimes\cwgen{E}_\alpha^+)
\>

and

\[
\rmat_H = \prod_{n=0}^\infty\exp\left(\left(\frac{d}{dv_1}\gen{K}_{i,+}(v_1)\right)_m\otimes\left( C_{i,j}(T^{1/2})\gen{K}_{j,-}(v_2+2n +1)\right)_{-m-1}\right) .
\]

The expressions for $\rmat_\pm$ do not need much explanation, apart from the fact that the arrows indicate in which order the product is taken with respect to the root ordering, $a(\alpha)$ is some normalisation and $\Delta_+$ denotes a set of all positive roots of the Yangian, i.e. $\Delta_+ = \{\beta + n\delta|\beta \text{ positive root of } \gltt, \delta \text{ affine root }, n\in\mathbb Z \}$. We should also note that these formulae have only been conjectured but not proven in \cite{Khoroshkin:1994uk}, so we make no claim that they are correct universally. However, when evaluating these expressions on the fundamental evaluation representation we get the correct result indicating, but not proving that the expressions are indeed correct.

The most nontrivial part is $\rmat_H$, it needs several explanations as it is quite complicated looking, and generically for simple Lie algebras it does not even become simple for fundamental evaluation representations but results in complicated expressions involving gamma functions. First, we note that if we have a matrix $X=(X_{i,j})$, then by $X(q)$ we mean the matrix $([X_{ij}])$, with $[X_{ij}]$ denoting the q number $[x] = \frac{q^{x} - q^{x}}{q - q^{-1}}$. Now $C(q)$ is proportional to the inverse of the q-Cartan matrix $A(q)$ such that all entiries contain only integer powers of q, i.e. we have

\[
C(q) = l(q)A(q)^{-1}.
\]

For simple Lie algebras $l$ is proportional to the dual Coxeter number \cite{Khoroshkin:1994uk}. However, in the case of $\gltt$ or $\alg{psl}(2|2)$, the dual Coxeter number is zero. One way to obtain integer valued inverse q-Cartan matrices is of course by just multiplying by the determinant of the q-Cartan matrix. For the extended one of $\gltt$ we have

\[
\cartq{} = \begin{pmatrix}
q+q^{-1}&-1&0&0\\
-1&0&1&1\\
0&1&q+q^{-1}&0\\
0&1&0&0
\end{pmatrix},
\]

its determinant is $[2]^2$. However, it suffices to multiply the inverse by only $[2]$ to obtain the minimal integer valued inverse Cartan matrix

\[
C(q) := [2](\cartq{})^{-1} = \begin{pmatrix}
1&0&0&1\\
0&0&0&[2]\\
0&0&-1&1\\
1&[2]&1&0
\end{pmatrix}.
\]

This seems to be the right choice for the universal R-matrix, even though it would be nice to better understand where the right choice comes from.

The operator $T$ is defined as the shift operator acting on functions in $v$, i.e. $T^x f(v) = f(v+x)$, hence by $C(T^x)$ we mean the matrix $C(q)$ where we substituted $q$ by the operator $T^x$.

Furthermore, the generators $\gen{K}_{i,\pm}(v)$ are defined as

\[
\gen{K}_{i,\pm}(v) = \log \gen{H}_{i,\pm}(v), 
\]

where 

\[
 \gen{H}_{i,+}(v) = 1 + \sum_{n=0}^\infty\csgen{H}_{i,n}v^{-n-1}, \quad \gen{H}_{i,-}(v) = 1 - \sum_{n=-1}^{-\infty}\csgen{H}_{i,n}v^{-n-1}.
\]

We will not explicitly write down the relations for the generators of negative degree, they can be derived in the same way as for simple Lie algebras as outlined in \cite{Khoroshkin:1994uk}. If we write $\left(\gen{K}(v)\right)_n$ we mean the n'th coefficient in the expansion in $v$ or $v^{-1}$ respectively.

\section{The R-matrix on the fundamental representation}\label{sec:funrep}

\subsection{Fundamental Evaluation Representation of $\Ya{\gltt}$ }

In this section we want to study the four dimensional fundamental evaluation representation of the Yangian $\Ya{\gltt}$ in Drinfeld's second realization. For the centrally extended case $\Ya{\psucentral}$ this was done in \cite{Spill:2008tp}, but in the purely ferminonic Chevalley-Serre basis, whereas here we use the distinguished one. We also have to supplement the information for the automorphism.

Let us first look at the presentation of the Yangian in the first realisation, i.e. using a linear basis such as the one defined in \eqref{def:linbase} for $\gltt$ \footnote{For more details about the relation between first and second realisation see e.g. \cite{Drinfeld:1987sy,Chari:1994pz, Spill:2008tp}.}. If we denote such linear basis by $\gen{J}^a_n$, $a=1,\dots 16, n=0,1,\dots$, then the fundamental evaluation representation of the Yangian in the first realisation is given by

\[\label{def:repfirst}
\genY{J}^a = u\gen{J}^a ,
\]

where on the right hand side we denote the generator $\gen{J}^a$ of $\gltt$ in the fundamental representation by the same symbol as the abstract Lie algebra generator. That means all generators of degree one are represented by the corresponding degree zero generator times the spectral parameter $u$. If we go to the generators of the Yangian in the second realisation the spectral parameter gets shifted in a different way for different generators. To derive this shift one evaluates the isomorphism between the first and second realisation on the fundamental representation. This isomorphism is given by

\begin{align}
\label{def:isom}
&\csgen{H}_{i,0}=\gen{H}_i,\quad \csgen{E}^+_{i,0}=\gen{E}_i,\quad \csgen{E}^-_{i,0}=\gen{F}_i,\nonumber\\
&\csgen{H}_{i,1}=\hat{\gen{H}}_i-v_i,\quad \csgen{E}^+_{i,1}=\hat{\gen{E}}_i-w_i,\quad \csgen{E}^-_{i,1}=\hat{\gen{F}}_i-z_i,
\end{align}
where the special elements are given by

\<\label{def:special}
v_i &=& \frac{1}{4}\sum_{\beta}\sprod{\alpha_i}{\beta}\acomm{\cwgen{E}^-_{\beta}}{\cwgen{E}^+_{\beta}} - \frac{1}{2}\csgen{H}_{i}^2, \nonumber\\ 
w_i &=& \frac{1}{4}\sum_{\beta}(-1)^{\beta i}\acomm{\cwgen{E}^-_{\beta}}{\comm{\csgen{E}^+_{i}}{\cwgen{E}^+_{\beta}}} - \frac{1}{4}\acomm{\csgen{H}_{i}}{\csgen{E}^+_{i}}, \nonumber\\
z_i &=& -\frac{1}{4}\sum_{\beta}\acomm{\comm{\csgen{E}^-_{i}}{\cwgen{E}^-_{\beta}}}{\cwgen{E}^+_{\beta}} - \frac{1}{4}\acomm{\csgen{H}_{i}}{\csgen{E}^-_{i}}.
\>

Evaluating on the fundamental evaluation representation of $\gltt$, we obtain

\begin{align}
v_i = a_i \csgen{H}_i, \nonumber\\
w_i = a_i \csgen{E}^+_i, \nonumber\\
z_i = a_i \csgen{E}^-_i.
\end{align}

with $a_i = (-\frac{1}{2},-1,-\frac{1}{2})$.

Now if we represent the generators of the first realisation as in \eqref{def:repfirst}, i.e. $\genY{J}^a = u\gen{J}^a$,  then we get the following representation for the generators of the second realisation:

\begin{align}
\csgen{X}_{i,1} = \tilde u_i\csgen{X}_i, \qquad \csgen{X} = \csgen{H},\csgen{E}^\pm ,
\end{align}

with $\tilde u_i = (u+\frac{1}{2},u+1,u+\frac{1}{2})$. 

So far we have not specified the isomorphism for the automorphism $\csgen{H}_4$. As there is no corresponding root vector $\alpha_4$, we first have to make clear that when we write $\sprod{\alpha_4}{\beta}$ in the expression for $v_4$, we mean the Killing form $\str(\csgen{H}_4,\csgen{H}_\beta)$. Then we can indeed consider the expression \ref{def:special} also for $v_4$, but when we evaluate it on the fundamental representation we find that $v_4$ is proportional to $\csgen{H}_4$ only up to a shift by the indentity matrix. This is not surprising as $\csgen{H}_4$ cannot be obtained as a commutator of any element in $\alg{gl}(2|2)$, and if we shift it by the central element we remain with the same commutation relations. Furthermore, the defining relations for the second realization are violated if we use this naive isomorphism, hence we define 

\[
 \csgen{H}_{4,1} := \genY{H}_4 - v_4 - \frac{1}{4}\gen{C},
\]

and find that all defining relations are satisfied. Furthermore, on the fundamental representation we obtain

\[
 \csgen{H}_{4,1} = \tilde u_4 \csgen{H}_4 = (u+1) \csgen{H}_4 .
\]

Having established the evaluation representation for the Chevalley-Serre generators, we can straightforwardly obtain the representation for all the other remaining generators. We can define the generators such that

\begin{align}
 \cwgen{E}^\pm_{bose,n} = \left (u+\frac{1}{2}\right )^n \cwgen{E}^\pm_{bose,0}, \qquad\cwgen{E}^\pm_{fermi,n} = (u+1)^n \cwgen{E}^\pm_{fermi,0},\qquad  \csgen{H}_{i,n} = u_i^n\csgen{H}_{i,0}  .
\end{align}

\subsection{Yangs R-matrix for $\gltt$}

Having established the fundamental evaluation representation of $\Ya{\gltt}$ it is straightforward to write down the representation of the universal R-matrix on the tensor product of two fundamental representations with spectral parameters $u, v$. First we note that the root parts of the universal R-matrix can be factorised as follows:

\<
\rmat_+ = \prod_{k=1,\dots 6}^\rightarrow\exp(-\sum^\infty_{n=0}\cwgen{E}_{\alpha_k,n}^+\otimes\cwgen{E}_{\alpha_k,-n-1}^-) ,\nonumber\\
\rmat_- = \prod_{k=6,\dots 1}^\leftarrow\exp(-\sum^\infty_{n=0}\cwgen{E}_{\alpha_k,n}^-\otimes\cwgen{E}_{\alpha_k,-n-1}^+) .
\>

Using the above evaluation representation we can sum up the resulting geometric series, and, using the fact that on the fundamental representation all roots are nilpotent, we obtain

\<
\rmat_+ = \prod_{k=1,\dots 6}^\rightarrow\exp(\frac{1}{u-v}\cwgen{E}_{\alpha_k}^+\otimes\cwgen{E}_{\alpha_k}^-) = \prod_{k=1,\dots 6}^\rightarrow(1 + \frac{1}{u-v}\cwgen{E}_{\alpha_k}^+\otimes\cwgen{E}_{\alpha_k}^-) ,\nonumber\\
\rmat_- = \prod_{k=1,\dots 6}^\rightarrow\exp(\frac{1}{u-v}\cwgen{E}_{\alpha_k}^-\otimes\cwgen{E}_{\alpha_k}^+) =\prod_{k=6,\dots 1}^\leftarrow(1+\frac{1}{u-v}\cwgen{E}_{\alpha_k}^-\otimes\cwgen{E}_{\alpha_k}^+).
\>

For the Cartan part we note that on the fundamental representation the generating functions sum up as 

\<
\gen{H}_{i,+}(\lambda) =  1 + \frac{1}{\lambda - u_i}\csgen{H}_i,\nonumber\\
\gen{H}_{i,-}(\mu) = 1 - \frac{1}{v_i - \mu}\csgen{H}_i .
\>

Note that $\gen{H}_{i,\pm}$ formally represent the same function, once expanded about infinity and once about zero. 
The terms appearing in $\rmat_H$ are the coefficients of the series

\<
\gen{K}_{i,+}(\lambda)' = \log \gen{H}_{i,+}(\lambda)' = \frac{1}{\lambda - u_i - \csgen{H}_i} - \frac{1}{\lambda - u_i} \nonumber\\
= \sum_{n=0}^\infty\lambda^{-n-1}\left((u_i - \csgen{H}_i)^n - u_i^n\right),\nonumber\\
\gen{K}_{i,-}(\lambda) = \log\gen{H}_{i,-}(\lambda) = \log \frac{u_i - \csgen{H}_i}{u_i}\frac{\frac{\lambda}{u_i-\csgen{H}_i} - 1}{\frac{\lambda}{u_i} - 1} = \nonumber\\
\log \frac{u_i - \csgen{H}_i}{u_i} + \sum_{n=1}^\infty\frac{\left((\frac{\lambda}{u_i})^n -(\frac{\lambda}{u_i-\csgen{H}_i})^n\right)}{n} .
\>

In $\rmat_H$ this leads to terms of the form

\<
 \prod_{m=1}^\infty\exp\left(-\sum_{n=0}^\infty (\frac{u - x}{v - y - 2n})^m\frac{1}{m}\right) &=& \prod_{n=0}^\infty(1 - \frac{u - x}{v - y - 2n})\nonumber\\
 = \prod_{n=0}^\infty (\frac{v - u - 2n - x}{v - y - 2n}) &=& \frac{\Gamma\left(\frac{u - v + y -x}{2}\right)}{\Gamma\left(\frac{y - v}{2}\right)} .
\>

Evaluating $\rmat_H$ completely using Mathematica we obtain the result that all gamma functions cancel, the full R-matrix on the fundamental representation is given by

\[\label{eqn:YangsR}
 \rmat = \frac{1+2(u-v)}{1-2(u-v)}\left(\frac{u-v}{u-v+1} + \frac{1}{u-v+1}\perm\right).
\]

Here, $\perm$ is the graded permutation operator. 
Note that to obtain the physical S-matrix for $\mathcal N=4$ SYM or Chern-Simons theory leading to the appropriate Bethe Ans\"atze the hermitian representation for the Yangian of $\utt$ requires $u \rightarrow i u$, with $u$ being the real valued rapidity.

\section{Discussion}

In this paper we derived the universal R-matrix for the double Yangian of $\gltt$. We also obtained the fundamental evaluation representation and evaluated the universal R-matrix on the tensor product of two fundamental representations, the resulting R-matrix is, as expected for R-matrices of $\alg{sl}(n), \alg{gl}(n|m)$ type, proportional to Yangs R-matrix $\left(\frac{u}{u+1} + \frac{1}{u+1}\perm\right)$. However, unlike for $\alg{sl}(n)$ considered in \cite{Khoroshkin:1994uk}, the gamma functions appearing in the Cartan part $\rmat_H$ of the R-matrix cancel, and the resulting R-matrix is a rational function depending on the difference of the spectral parameters only. This is required to obtain the correct Bethe Ansatz for weakly coupled $\mathcal N = 4$ SYM as well as Chern-Simons theories. However, the factor we obtain seems to be not exactly correct; this could be explained by a change of basis and additional diagonal twists. One can obtain this S-matrix by considering the weak coupling limit of the full S-matrix of \cite{Beisert:2005tm} leading to the full long-range Bethe equations of $\mathcal N = 4$ SYM. A natural question following the construction of the universal R-matrix for $\gltt$ is if one can also derive the universal R-matrix leading to the all loop S-matrix. It was shown that this all loop S-matrix is also invariant under Yangian symmetry \cite{Beisert:2007ds}, and this Yangian can also be formulated in the second realisation \cite{Spill:2008tp}. Then one is lead to the problem how to treat the additional central elements which are necessary to describe the long-range effects. On the level of the classical r-matrix \cite{Beisert:2007ds} both at strong and weak coupling it was shown how to effectively remove these extra central elements and the braiding describing length changing \cite{Plefka:2006ze} and obtain the classical r-matrix from a classical double of a deformed $\gltt$ loop algebra. This work makes it plausible that the quantum double and universal R-matrix should also exist for the centrally extended algebra $\psucentral$ and will be a deformation of the universal R-matrix for $\mathcal Y(\gltt)$ considered here. Note that one can rewrite the S-matrix of \cite{Beisert:2005tm} in such a way which makes it apparant which part comes from the Yangian structure and which part comes from the central extension and braiding \cite{Torrielli:2008wi}. Finding the universal R-matrix is an important task as this universal form might be used to compute bound state S matrices as done on the level of representations in \cite{Arutyunov:2008zt} or give some ideas about the origin of the dressing phase. Another important task is finding out if the S-matrix of $\mathcal N = 6$ Chern-Simons theory is really the same as the one for $\mathcal N = 4$ SYM, only that for the former one has two different kinds of particles, whereas for the latter one needs to take the tensor product of two fundamental representations to form a magnon. In \cite{Ahn:2008aa} it was argued that this is the case as this S-matrix leads to the Bethe equations \cite{Gromov:2008qe}. However, one should thouroughly derive the central extension and the braiding of \cite{Plefka:2006ze} as this is required to obtain a non-trivial S-matrix not depending on the difference of spectral parameters only. Furthermore, one should find Yangian symmetry also on the Chern-Simons side as done for $\mathcal N = 4$ for tree level \cite{Dolan:2003uh, Dolan:2004ps}.

Some other relating questions following this work concern mathematical generalisations. It is easy to convince oneself that one can generalise the methods used here to derive the universal R matrices for all Yangians based on $\alg{gl}(n|m)$. It would be interesting to find out if the prefactors also have such simple structure as in the case of $\gltt$, and if the formulae will still be practical in the sense that one can obtain Yangs R-matrix on the fundamental representation straightforwardly by plugging in the represented generators. Furthermore, one should study more complicated representations of the Yangian. Other interesting generalisations would involve studying q deformations as done on the level of the spin chain and S-matrix in \cite{Beisert:2008tw} or studying more involved algebras such as the related exceptional superalgebra $\alg{d}(2,1;\alpha)$ \cite{Heckenberger:2007ry}.

\section*{Acknowledgements}

I want to thank P. Koroteev, $\Omega$ Mekareeya, A. Rej and A. Tseytlin for useful comments on the manuscript. This research was supported in part by DARPA and AFOSR through the grant FA9550-07-1-0543 and by the National Science Foundation under Grant No. PHY05-51164, when the author was visiting the KITP, Santa Barbara. The author is grateful to the Deutsche Telekom Stiftung for supporting him with a PhD fellowship.

%\addcontentsline{toc}{chapter}{Bibliography}
%\bibliographystyle{amsalpha}
%\bibliographystyle{unsrt}
\bibliographystyle{myutphys}
\bibliography{universalr}
\end{document}